\documentclass[acmsmall,authorversion]{acmart}
\AtBeginDocument{%
  \providecommand\BibTeX{{%
    \normalfont B\kern-0.5em{\scshape i\kern-0.25em b}\kern-0.8em\TeX}}}

\usepackage {threeparttable}
\usepackage {pifont}
\begin{document}

\title{Octopus: A Heterogeneous In-network Computing Accelerator Enabling Deep Learning for network.}

\author{Dong Wen}
\email{wendong19@nudt.edu.cn}
\affiliation{%
  \institution{National University of Defense Technology}
  \country{China}
}

\author{Tao Li}
\email{taoli@nudt.edu.cn}
\affiliation{%
  \institution{National University of Defense Technology}
  \country{China}
}

\author{Chenglong Li}
\email{lichenglong17@nudt.edu.cn}
\affiliation{%
  \institution{National University of Defense Technology}
  \country{China}
}

\author{Pengye Xia}
\email{pyxia@nudt.edu.cn}
\affiliation{%
  \institution{National University of Defense Technology}
  \country{China}
}

\author{Zhigang Sun}
\email{sunzhigang@nudt.edu.cn}
\affiliation{%
  \institution{National University of Defense Technology}
  \country{China}
}








\renewcommand{\shortauthors}{Trovato and Tobin, et al.}

\begin{abstract}
Deep learning (DL) for network models have achieved excellent performance in the field and 
are becoming a promising component in future intelligent network system.
Programmable in-network computing device has great potential to deploy DL for network models, 
however, existing device cannot afford to run a DL model.
The main challenges of data-plane supporting DL-based network models lie in computing power, task granularity, model generality and feature extracting.

To address above problems, we propose Octopus: a heterogeneous in-network computing accelerator enabling DL for network models. 
A feature extractor is designed for fast and efficient feature extracting. 
Vector accelerator and systolic array work in a heterogeneous collaborative way, offering low-latency-high-throughput general computing ability for packet-and-flow-based tasks. 
Octopus also contains on-chip memory fabric for storage and connecting, and Risc-V core for global controlling. 
The proposed Octopus accelerator design is implemented on FPGA.
Functionality and performance of Octopus are validated in several use-cases, achieving performance of 31Mpkt/s feature extracting, 207ns packet-based computing latency, and 90kflow/s flow-based computing throughput. 

\end{abstract}




\maketitle

\section{Introduction}
Various emerging deep learning (DL) models for network have been investigated and applied
in many key network functions. 
DL for network models achieve excellent performance in the field of network traffic analysis \cite{zeroday, survey0, payloadcnn0, payloadcnn1, payloadcnn2, 1dcnn0, 1dcnn1, 1dcnn2}, 
intrusion detection \cite{usecase1, zeroday, survey0, sever2, iisy}, traffic routing\cite{routegnn0, routegnn1, routegnn2, routegnn3,routegnn4}, traffic control \cite{routing0, routing1, routing2} and other network functions.
Deep learning for network is becoming a promising component in future automated and intelligent network system. 

In-network computing based on programmable network devices has great potential to deploy DL for network models. 
Compared with network systems implemented by high-performance servers \cite{sever0, sever1, sever2}, programmable in-network computing devices save network-to-host data-moving overhead, process traffic in real-time and support feature extracting with low latency. 
In a word, programmable-device-based in-network computing has fast access to traffic information, which is necessary for low-delay and high-throughput DL for network models. 
P4 \cite{p4, fast, menshen, rmt} and many-core-based \cite{spin, pspin} data-plane are prevailing device for in-network computing in current practice.

We ask questions: \textit{To make data-plane enable in-network computing for DL models, what is the real challenge?}
Based on our observation, four challenges are most important: 
(i) \textit{Computing power}. 
Computing power is the most important requirement of deploying a DL model, because DL models demand larger computation than traditional strategies.
However, considering the low-latency-high-throughput characteristic of network traffic, it is not straightforward to simply insert an existed accelerator into the data-plane and satisfy these constraints.
(ii) \textit{Task granularity}. A data-plane should support both packet-based and flow-based DL for network models. Both of them are common and important in real network environment \cite{routing2, survey0, routegnn1}. 
However, two granularity poses significantly different requirements on storage, computing latency and throughput for hardware.
(iii) \textit{Model generality}. A simple specific architecture can only accelerate a certain kind of models. To fully enable common DL for network models such as convolution neural networks (CNN), multi-layer perceptron (MLP) and transformers, a generality architecture is needed. 
(iv) \textit{Feature extracting}. 
Feature extracting is an important step to run a DL-based network models because traffic feature is the input of deep neural networks (DNN). 
Feature extractor should keep balance between hardware resource occupation and extracting speed.
The detail illustration will appear in next chapter.

As common platforms for in-network computing, P4 and many-core-based data-plane cannot afford to run a DL for network model. 
Their architectures lack hardware computing units for DL models.
Secondly, their traffic feature extraction is not efficient. P4 hardware occupy many stages of pipeline to extract feature, leading to waste of hardware resource. Many-core-based devices depends on CPU programs, which is time-consuming.
To enhance the computing power of data-plane, Tarurs \cite{tarurs} argues to improve P4 hardware with computing accelerators.  
Tarurs basically adds a specific hardware engine for CNN between ingress and engress pipelines of a P4 device and achieve DL inference delay within hundreds of nano-seconds (ns). 
However, Tarurs only focuses on packet-based models and ignores tasks based on flow.
Its generality is also limited by pipeline-structure accelerator, which is difficult to deploy other common models in the field.
To conclude, existing data-plane architectures are unable to support DL-based network models, and the method of straightforwardly inserting an accelerator into data-plane cannot address above challenges well.
Thus it is non-trivial to design a new accelerator for in-network DL computing.

\begin{figure}[h]
  \centering
  \includegraphics[width=\linewidth]{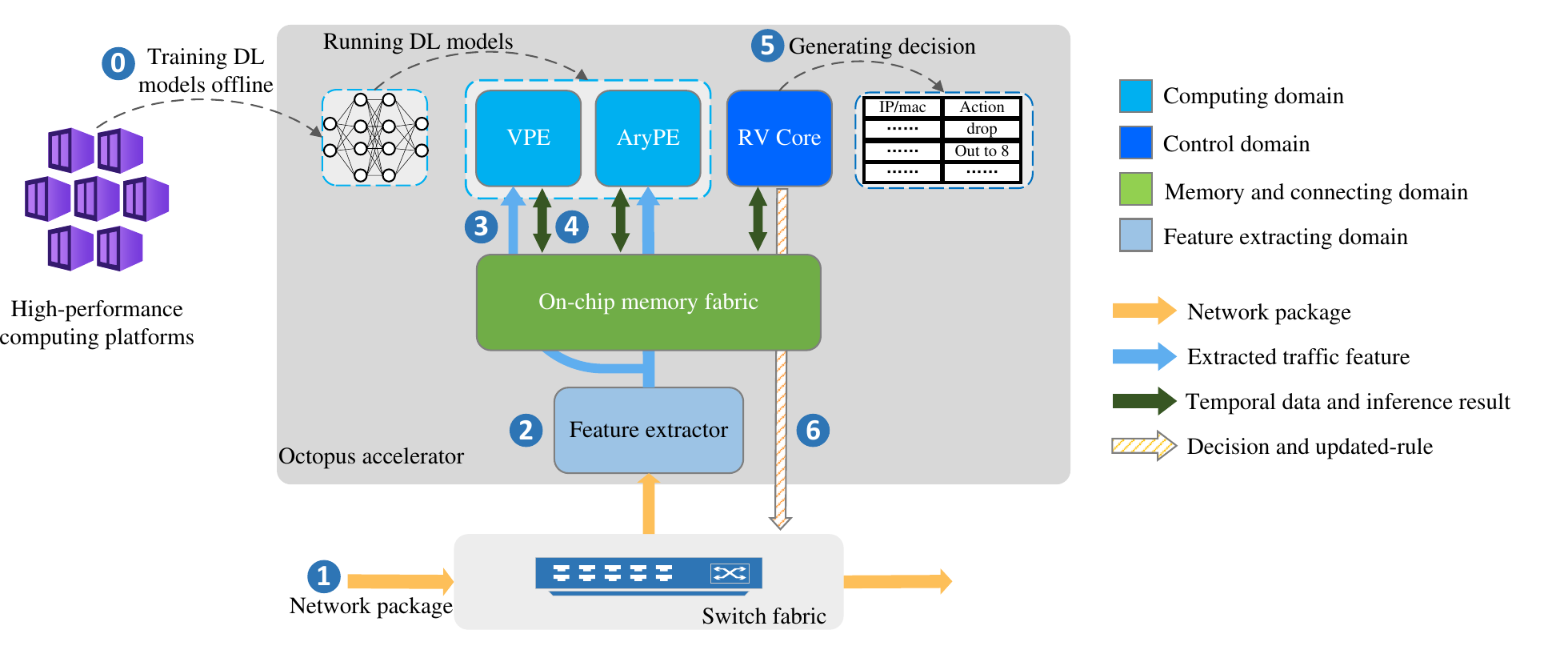}
  \caption{The overview of Octopus}
  \Description{overview}
  \label{overview}
\end{figure}

In this paper, we propose Octopus: a heterogeneous in-network computing accelerator for DL-based network models. 
Figure \ref{overview} shows the overview of Octopus architecture and its working procedure. 
The proposed architecture is divided into four function domains. 
Feature extracting domain first receives packet from switch fabric, then feature extractor rapidly extracts a small set of meta-feature, then derives other high-level feature under configuration. 
Fast feature extracting with high hardware efficiency are achieved in our design.
Acquired traffic feature is stored in memory and connecting domain, and sent to the computing domain afterwards.
Computing domain includes vector process element (VPE) and array process element (AryPE), providing computing power for DL models. 
VPE adopts single-instruction-multiple-data (SIMD) and very-long-instruction-word (VLIW) structure, offering forwarding-comparable computing latency for packet-based tasks.
AryPE is consisted of systolic array \cite{array}, a throughput-optimal design for matrix multiplication which provides higher throughout for flow-based tasks.
In Octopus, VPE and AryPE are not simply placed together on chip.
With on-chip memory fabric functioning as data-exchanging path, VPE and AryPE work in a heterogeneous collaborative computing way.
VPE provides flexible and parallel vector functions, benefiting computing efficiency of whole architecture.
To realize model generality, the architectures of VPE and AryPE possess general linear-algebra arithmetic ability and non-pipeline structure. 
Also, a simple instruction set of computing domain is proposed to help users program their DL models.
Risc-V (RV) core in control domain is in charge of loading configuration and instructions for other function modules, generating controlling signals and translating DL inference results into decisions for switch fabric.

A brief working procedure of Octopus is as below: 
first, training DL models on an offline platform (\textcircled{0}) is before the start of working.
When a pre-trained model is deployed, arriving network packets (\textcircled{1}) are uploaded to Octopus. 
Traffic feature is extracted (\textcircled{2}),
and sent to computing domain for DL models inference (\textcircled{3}). On-chip heterogeneous collaborating happens between VPE, AryPE and on-chip memory fabric (\textcircled{4}). 
RV core generates decisions based on computing results (\textcircled{5}),  
and finally, latest rules are sent to switch device for updating (\textcircled{6}).


To best of our knowledge, Octopus is the first heterogeneous in-network accelerator sufficiently supports DL for network models. 
In nut-shell, our contributions are following: 

(1) We investigate and clarify four key challenges of making data-plane enable in-network computing for DL models.
We believe such principles are instructive for improving and designing future intelligent network programmable device.

(2) To address above challenges, we propose Octopus, a heterogeneous in-network accelerator that sufficiently enables DL for network models. 
The architecture of Octopus satisfies the demand of DL models on computing power, task granularity, model generality and feature extracting.

(3) A novel heterogeneous architecture is designed in Octopus: 
a fast and hardware-efficient feature extractor, a heterogeneous collaborative computing domain, an on-chip memory fabric for storage and connecting, and a RV core controller generating decisions. 


(4) We implement Octopus on FPGA, with performance of 31Mpkt/s feature extracting, 207ns packet-based inference delay and up to 90kflow/s flow-based inference throughput. 
Several use-cases are arranged, validating the functionality and performance of Octopus.


\section{Background and Motivation}

\subsection {DL for network models}
Inspired by the success in computer vision (CV) \cite{vgg, resnet, vit} and nature language process (NLP) \cite{attn}, advanced DL models like CNN \cite{payloadcnn0, payloadcnn1, payloadcnn2, 1dcnn0, 1dcnn2}, MLP \cite{usecase1, mlp0, mlp1, mlpmix} and transformers \cite{usecase3, netattn0, payloadattn, payloadattn1} are introduced to network functions. 
The insight of DL-based network models is training a DNN to explore the hidden information of network traffics and predict network behaviours. The training procedure is completed on collected data-set and high-performance CPU/GPU in an offline way. 
When deployed in real world network environments, network managers only need to feed pre-trained DL models with real time traffic feature and start online inference.

\begin{figure}[h]
  \centering
  \includegraphics[width=\linewidth]{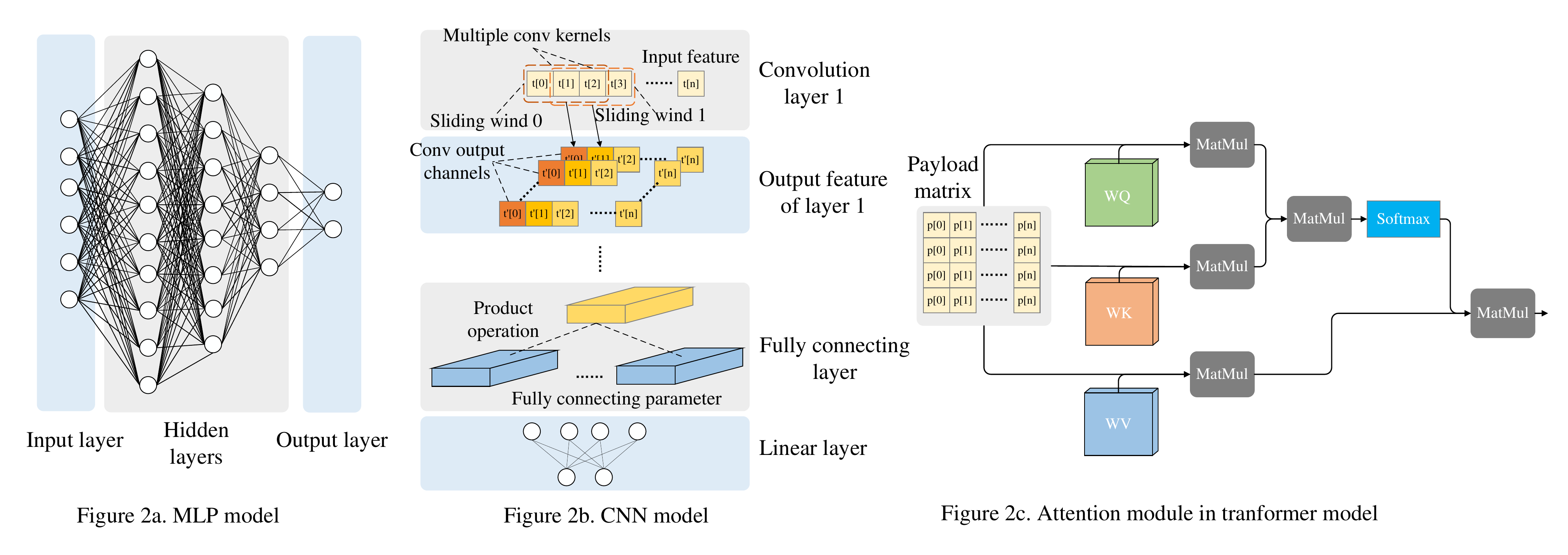}
  \caption{Three typical DL models for network}
  \Description{dl model}
  \label{dl model}
\end{figure}

Figure \ref{dl model} shows three typical DL models used in network field. 
MLP model \cite{mlptheory} is shown in figure \ref{dl model}a, which is applied in both packet-based or flow-based tasks.
MLP takes diverse feature as input, like flow duration time, flow size and TCP/UDP flag in \cite{usecase1, mlp0, mlp1}.
Feature is sent to hidden layers from input layer.
Hidden layers contain multi-layer neuron units and non-linear activation units.
A hidden layer with $a$ inputs and $b$ outputs can map to a parameter matrix with dimension of $(a, b)$.
During inference, vector or matrix multiplication is performed between input feature and parameter matrix.
Output layer gives final prediction, for example, whether in-coming packet or flow is malicious traffic.
One-dimensional (1D) CNN models in figure \ref{dl model}b, however, receive time series data as input.
The input feature can be an n-dimension ordered vector of packet size, packet arrival time or other features of top-n packets in a flow. 
The convolution kernels in CNN extracts high-level information of neighbour elements in sliding windows. 
The convolution operation can be transformed into standard matrix multiplications via img2col function \cite{torch}.
After convolution layers follow fully connecting layer and linear layers for final prediction. 
1D-CNN models are widely used in flow-level traffic classification tasks \cite{zeroday, survey0,1dcnn0, 1dcnn1, 1dcnn2}. 
Transformer in figure \ref{dl model}c is the state-of-the-art models in the NLP field and introduced in network tasks . 
Different from aforementioned models, transformers in \cite{usecase3, payloadattn, payloadattn1, netattn0} choose payload slice from top-k packets as input feature. 
Multi-head self-attention is the core mechanism in transformer model. 
To find long-distance information from payload, attention mechanism applies three modules : \textit{Q}, \textit{K}, \textit{V}, which is represented by parameter matrix \textit{WQ}, \textit{WK}, \textit{WV}, respectively.
Top-n bytes of payload from top-k packets in a flow is organized into a payload matrix, then series of matrix multiplications are performed between payload and parameter, as depicted in figure \ref{dl model}c.
Softmax is the non-linear activation function in transformer.
The output of attention mechanism is then sent to following layers for final prediction. 
Although deployments do not need rounds of training, the inference of above models is still time-consuming and needs in-network DL accelerators. 

\subsection {DL accelerators}
There are plenty works focusing on DL accelerator in CV and NLP cases \cite{eie, me, cnnaccl0, cnnaccl1, vecaccl1}, however, DL accelerators for in-network computing have not been thoroughly investigated. 
In-network computing requires low-latency-high-throughput devices to match traffic forwarding speed and throughput performance, posing new challenge for existing DL accelerators.
Existing DL accelerators can be roughly divided into three types: specific accelerator\cite{eie, me, cnnaccl0, cnnaccl1}, vector accelerator \cite{vecaccl, vecaccl1} and systolic array. 
Specific accelerators are not discussed in this paper, for they are designed for certain kinds of DL model thus cannot satisfy generality requirement. 
Vector accelerator and systolic array can be seen as general architectures for DL models because as illustrated above, computation of different DL models can be transformed into general linear-algebra arithmetic tasks like matrix multiplication.

\begin{figure}[h]
  \centering
  \includegraphics[width=\linewidth]{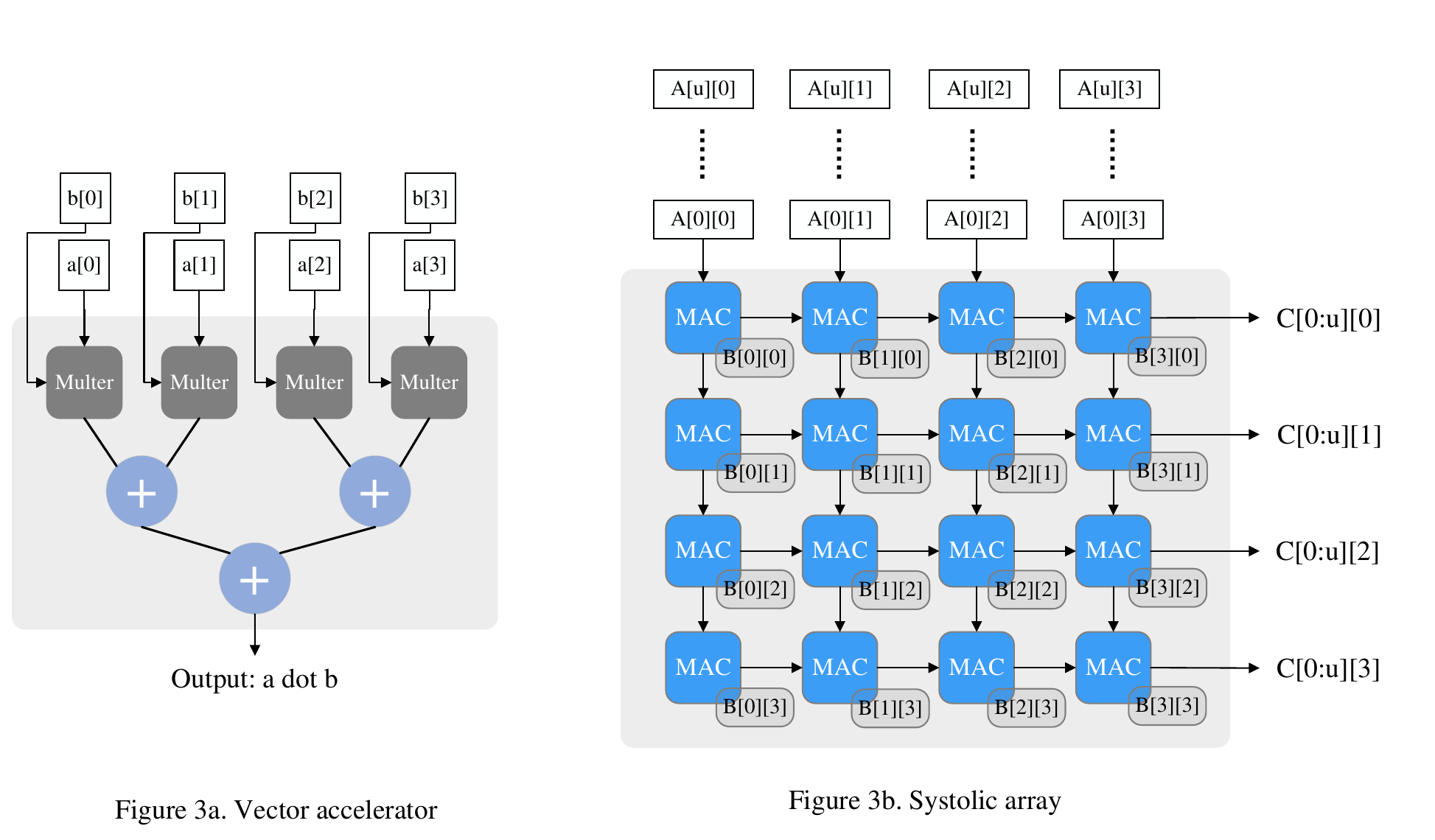}
  \caption{The architecture of vector accelerator and systolic array}
  \Description{accl_arch}
  \label{accl_arch}
\end{figure}

Vector accelerator in figure \ref{accl_arch}a. focuses on general vector product. The architecture of vector accelerators usually contains parallel multiplying units and tree-shape adders. This design can provide the delay of $O($log$(\frac{m}{2})/n)$ and throughput of $O(n/$log$(\frac{m}{2}))$, $m$ represents the number of computing units within one accelerator, and $n$ is the number of accelerators.
Systolic array in figure \ref{accl_arch}b is firstly proposed by \cite{array} and is widely used in high-end GPU and TPU. 
Systolic array is designed for streaming general matrix multiplication and proved to have the theoretical optimal throughput. 
Its delay is of $O(\sqrt{m}/n)$ and throughput is $O(n)$. 
In practice, however, the computing efficiency of systolic array is limited by under-utilization and block matrix multiplication problems \cite{array0}.
Taking the computation of CNN as an example: 
under-utilization mainly occurs in shallow layers of CNN, where the dimension of feature and parameter are too small to fill all multiplication-accumulating (MAC) units in systolic array, harming hardware utilization.
Block matrix multiplication happens in deeper layers of CNN, with the expanding of feature and parameter size, the dimension of computation exceeds the size of systolic array and has to conduct block matrix multiplication.
In this case, systolic array has to interrupt for block matrix aggregations, lowering computing efficiency. 
Such efficiency loss is common in CV or NLP cases, and to some extent, such loss seems to be acceptable in these tasks \cite{TPU}. 
However, in-network computing raises stringent requirements on delay and throughput for accelerator, and might not tolerate such efficiency loss.

In conclusion, two architectures show significant difference on computing delay and throughput, but neither of them meets the delay-and-throughput requirement of in-network computing.
At first glance, simply combining two accelerators appears to be a way for addressing challenges, but it still cannot solve the efficiency-loss problems.

\subsection{Analysis on key challenges}
In this section, we illustrate four challenges of deploying DL for network models by in-network computing device.

\textit{\textbf{Computing power}} is a basic precondition for running a DL model. 
As discussed before, in-network computing for DL models should be both latency-sensitive and throughput-demanding.
However, existing general DL accelerators cannot meet all requirements.
Vector architecture owns better computing latency but lower computing throughput, while systolic array is better on throughput but with longer delay. 
The computing resource of in-network computing device is limited by chip area and power, thus scaling only one types of accelerator to meet performance requirement is in-optimal . 
It is reasonable to take both architectures for low latency as well as high throughput.

In Octopus, VPE and AryPE are not straightforwardly combined, but works in a heterogeneous collaborative method. 
At first space, VPE takes charge of low-latency tasks, while AryPE targets for throughput-demanding computation.
To further address the efficiency-loss problems and pursue higher performance,
VPE supports AryPE with high-efficiency small-size operations and blocking matrix aggregation. 
A heterogeneous collaborative mode is formulated, enhancing overall computing ability.

\textit{\textbf{Task granularity}} poses significant different requirements on batch-size, storage, computing latency and computing throughput of hardware, as shown in table \ref{flow pkt}.
packet-based DL models demand hardware design with extreme-low delay of hundreds of nano-seconds(ns), for it being expected to match the forwarding speed of packets. 
Also, the batch-size of the model is limited by the number of PHY ports on device, because one PHY port receives at most one packet at a time.
From the line-speed of data-plane and expected delay, the throughput requirement of packet-based models can be calculated. 
Finally, packet-based DL for network models conduct inference per-packet, so they do not need history information and occupy no or little storage.

On contrast, flow-based DL models take more importance on throughput, because a data-plane device may track up to thousands of different flows. 
Conducting DL inference on these flows requires an accelerator of higher throughput and higher computing power.
In this case, the batch-size is decided by the scale of tracked flows.
Computing delay should be around several mile-seconds (ms) to hundreds of ms to provide periodical real-time decision for a flow, as \cite{survey0, zeroday, fenxi} argued. 
Besides, flow-based models depend on the history information of top-n packets in a flow, which requires hardware to provide sufficient storage for the information.
Models of two granularity are both important and common in real world task. It is important for in-network computing hardware to satisfy their different requirements. Octopus allocates packet-based tasks to VPE and flow-based ones for AryPE, meeting delay and throughput demands for two granularity. Also, on-chip memory fabric provides storage for flow information.

\textit{\textbf{Model generality}} 
requires hardware to support common DL models in network field. 
CNN\cite{zeroday, 1dcnn1, 1dcnn2, payloadcnn0, payloadcnn1, payloadcnn2}, MLP\cite{usecase1, mlp0, mlp1}, long-short term memory (LSTM) and other advanced DL models are widely used in traffic analysis, DBN and graph neural network (GNN) are adopted for routing tasks \cite{routegnn0, routegnn2, routegnn3, routegnn4}, transformer models \cite{netattn0, usecase3, payloadattn, payloadattn1} based on self-attention is one of the emerging choice in the field.
Based on our observation, We argue that CNN, MLP and transformer are more common choice in DL network models. In-network computing hardware should support these models at least.

The generality of Octopus is realized in architecture and instruction set design.
Compared with specific accelerator, 
VPE and AryPE can perform general linear-algebra arithmetic computing for DL models.
And most of all, both of them can be programmed by the proposed instruction, which bridges an interface between programmer and hardware.
With simple instructions like vector product and matrix multiplication in the instruction set, user-define DL models can be easily deployed on Octopus. 

\textit{\textbf{Feature extracting}} is an important step to run a DL for network models. 
As introduced in section 2.1, different models may adopts different traffic features as input.
According to our statistic, there exists over forty different features used in DL for network models. 
These features can be divided into several categories: size, time, speed, direction, payload and protocol features of a packet or a flow. 
All collected features are noted as "whole feature set" and shown in appendix.
Obviously it is uneconomic to extract all features because a DL model may only take a scratch of features, while hardware cost and extracting delay for all features are too high.

To simplify feature extracting, we propose "meta feature set" contrast to whole feature set. 
Meta feature set is an atomic set of feature which can derive whole feature set via simple configurations. 
Table \ref{meta set} shows all elements.
When a model is like \cite{usecase1} and needs flow duration time for example, it only needs to accumulate the pkt\_arv\_intv of all packets. 
The meaning of meta set is twofold: (i) the extracting of meta-set features is hardware-friendly; (ii) to support whole feature functionality, it demands configurable feature extractor design.
The feature extractor in Octopus is implemented by hardware logic with configuration ability, achieving a balance between hardware efficiency and fast extracting.

\begin{table}
  \caption{Difference between packet-based and flow-based DL for network models}
  \label{flow pkt}
  \begin{tabular}{cccccc}
    \toprule
    Granu.&Batch&Delay Req.&Througput Req.& Storage&Target\\
    \midrule
    pkt-based&1-10&hundreds of ns&line-speed with batch&no hist. info&per-pkt\\
    flow-based&1-1M&several ms&1-1M flow/ms&need hist. info&top-n pkt\\
    
  \bottomrule
\end{tabular}
\end{table}

\begin{table}
  \caption{Meta feature set}
  \label{meta set}
  \begin{tabular}{lll}
    \toprule
    Element &Meaning & Category\\
    \midrule
    pkt\_size&the size of packet & size\\
    payload&top-n Bytes of payload & payload\\
    pkt\_arv\_intv&the interval time of packet arrival & time\\
    dir&the direction of packet & direction\\
    tuple&IP tuple of packet & protocol\\
    flag&TCP/UDP/ICMP flag & protocol\\
  \bottomrule
\end{tabular}
\end{table}

\section{Architecture}
Figure \ref{overview} depicts the overview design of Octopus. The architecture is divided and introduced in four domains: feature extracting domain, computing domain, memory and connecting domain and control domain. 

\subsection{Feature extracting domain}
\begin{figure}[h]
  \centering
  \includegraphics[width=0.75\linewidth]{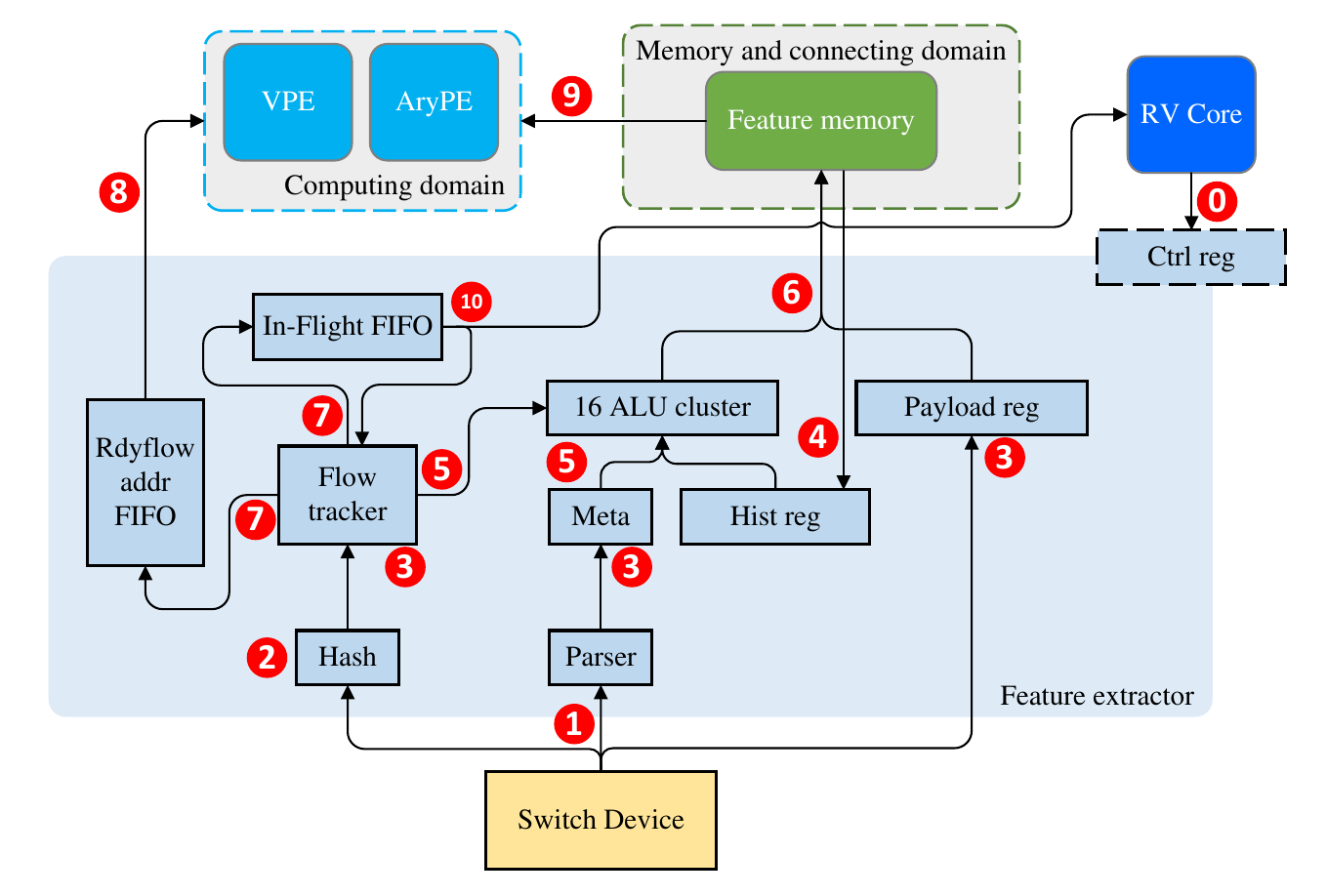}
  \caption{Feature extractor design}
  \label{fea_ext}
  \Description{Overview}
\end{figure}

Feature extractor in figure \ref{fea_ext} is the function component in this domain.
The input is packets from data-plane, and the output are for two parts: extracted feature to on-chip memory fabric, and the feature address to computing domain.


In initializing stage, RV core dispatches configuration information into control register firstly (\textcircled{0}).
And when a new packet arrives (\textcircled{1}), extractor sends its IP tuple to a hash module, generating an address for current tuple (\textcircled{2}). This address represents the position of an IP tuple (or a flow) in feature memory and flow tracker.
There is a storage element in flow tracker, designed for establishing state for new flows, keeping state information of existing flows, and freeing space for outdated flows. 
State information like MAC address, packet number of current flow, the timestamp of last packet are kept in the storage element.
When new hash address arrives, flow tracker reads the content in address, if packet number is zero, then it is judged as a new flow, or there comes a hit one (\textcircled{3}). 
Flow tracker updates latest information for each packet (\textcircled{5}). 
When it reaches a threshold, for example top-n packets of a flow has all arrived, flow tracker will push this address into ready-flow address FIFO for computing domain to fetch. 
Also, the address will be pushed into in-flight FIFO and the content in storage element is frozen (\textcircled{7}). 
Computing domain fetches feature address (\textcircled{8}) and loading data from feature memory (\textcircled{9}) for computing in order.
When computing is finished, FIN signal will be sent to feature extractor, so used flow information is no more needed, and storage space can be recycled.
Flow tracker will read out the top address in in-flight FIFO and set packet numbers in this address to zero. 
The address is also sent to RV core for decisions (\textcircled{10}).
Above is the establishment and freeing procedure of a flow in feature extractor.

The procedure of extracting meta set feature from packets is below: 
When receiving a new packet, packet header and timestamp is sent to parser (\textcircled{1}). 
Pkt\_size, tuple, and flag in meta set can be directly acquired in certain fields of header (\textcircled{3}), but to get pkt\_arv\_intv needs and update other flow features, history information in flow tracker and on-chip memory fabric are needed. 
ALU cluster is a configurable unit for feature extracting. 
The source of ALU cluster is meta register and history register, which keeps meta set feature of current packet and history information of the flow respectively. 
Meta register is with 13 Byte width and history register is with 16 Byte width.
The content in history register is pre-fetched from feature memory (\textcircled{4}).
If incoming packet belongs to a hit flow, flow tracker will send last timestamp for subtraction (\textcircled{5}), otherwise it will be set as zero, for current packet is the first packet of a new flow.
The output of ALU cluster is of 16 Byte width as well and is directly sent to feature memory (\textcircled{6}).
There are 16 ALUs in the cluster, each binds to one byte of output. 
An ALU can conduct micro operations like $add, subtract, max, min, wr$ on source data.
For example, flow duration time is kept in 0th byte of a feature memory word and pkt\_arv\_intv is in 7th byte of meta register, so the 0th ALU conducts $(add, \$0, \$7, \$0)$ operations, adding current packet arriving time to history time to update latest flow duration time.
The latest data is written back to 0th byte and then sent to feature memory.
The micro operations used by ALU clusters is stored in control register and loaded into each ALU when initialized. 

Payload in meta set is also a useful feature. In some works \cite{payloadattn, payloadattn1, payloadcnn0, payloadcnn1, payloadcnn2, usecase3}, DL models take top-k bytes of payload as input. Payload register in feature extractor truncates a certain length of payload (\textcircled{3}) and writes feature memory (\textcircled{6}). The implement and hardware resource utilization is evaluated in next chapter.

\subsection{Computing domain}
Computing domain includes VPE and AryPE, providing computing ability for DL-based network models. 
VPE is for latency-sensitive packet-based tasks and AryPE is for throughput-demanding flow-based tasks. 
Two computing elements also work in a heterogeneous collaborative method via on-chip memory fabric.

\subsubsection{VPE}
\begin{figure}[h]
  \centering
  \includegraphics[width=\linewidth]{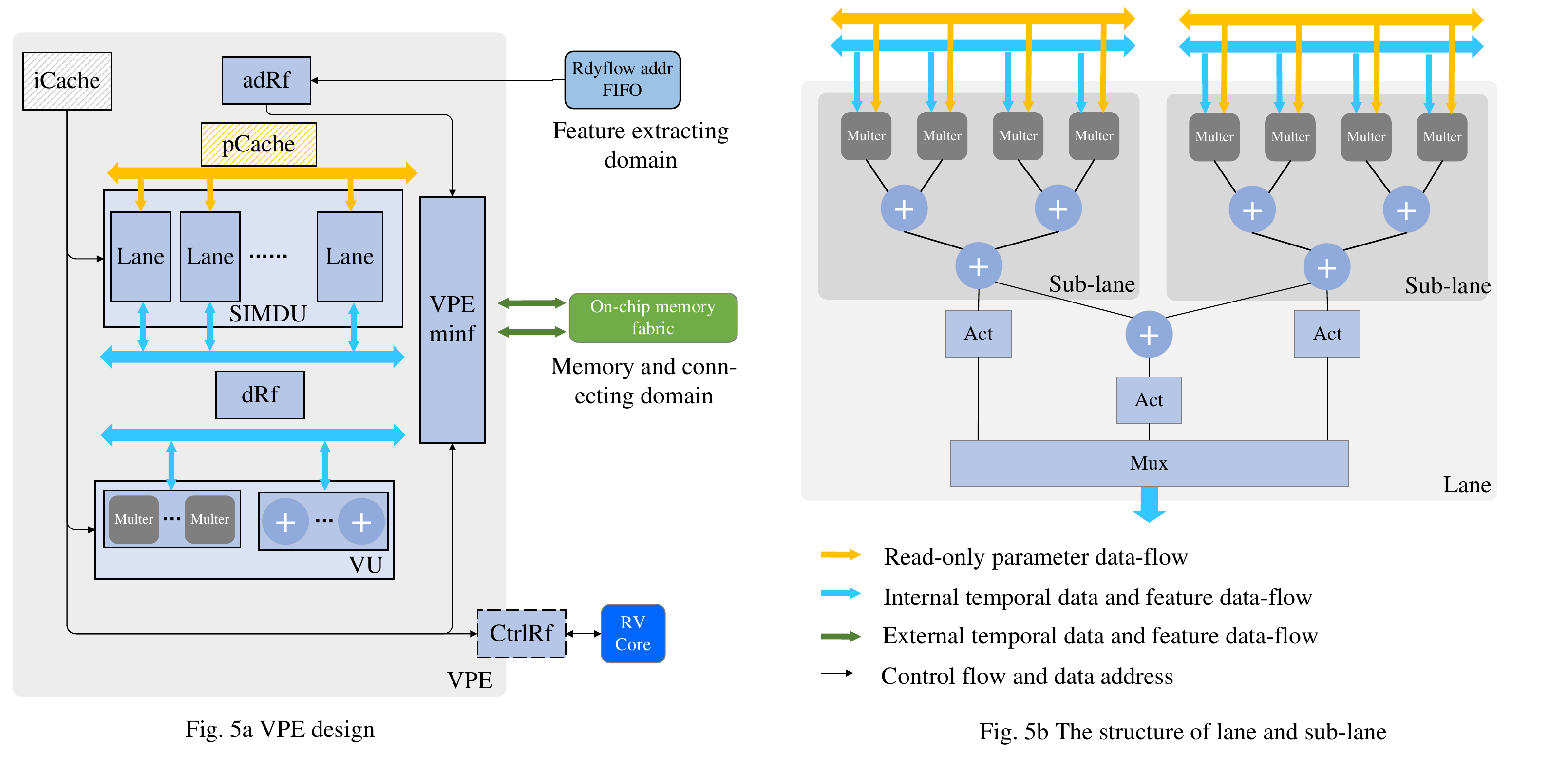}
  \caption{The hierarchy structure of VPE}
  \Description{Overview}
  \label{VPE}
\end{figure}
VPE is the low-latency computing unit for packet-based tasks in computing domain.
The architecture of VPE is detailed in figure \ref{VPE}, it includes four main function modules: SIMD unit (SIMDU), vector unit (VU), memory interface (Mif) and control register file (ctrlRF). 
All parts works under the control of VLIW. 
SIMDU is designed for low-latency vector product computation, and VU is for vector adding and element-wise multiplication of vector.
Data register file (dRf) and address register file (adRf) is shared by SIMDU and VU.
DRf is the space for temporal data registering and fast data exchange between two computing modules.
AdRf stores the address of source and destination data in on-chip memory fabric.
Mif moves data between VPE and memory fabric. 
Instruction cache (iCache) stores VLIW instructions, and DL model parameter cache (pCache) stores read-only parameter.
CtrlRf is the control interface between control domain and VPE.

\textit{SIMDU} is consisted of several basic computing units, noted as lane. A lane is consisted of two sub-lanes. 
In our SIMD design, all lanes in a SIMDU receive same instructions and different data (different slice of DL parameters, to be more specific), as shown in figure \ref{VPE}b. 
There are four multiplication units and an adder-tree in a sub-lane, together with an non-linear activation function stage.
One sub-lane can perform a four-element-width vector product plus an activation operation like ReLU or GeLU.
A lane is programmable to combine two sub-lanes for an eight-element-width vector production or separately use them for two smaller product. 
The reason for such design is that the size of CNN kernels in network applications is usually between 1-8, so a lane can finish two convolution operations with kernel size 1-4 or one convolution operation with kernel size 5-8, keeping balance between flexibility and performance.
The computing delay of SIMDU is decided by the depth of adder trees and the hardware delay of adder/multiplication units.

\textit{VU} contains parallel multiplication and adder units for element-wise multiplication and vector adding, which are common in normalization and blocking matrix multiplication in DL models.
VU is designed for offloading such operations to maximize the computing efficiency of SIMDU and AryPE. 

\textit{Mif} manages all memory access between VPE and on-chip memory fabric. 
Mif is designed with multiple memory channels, enabling VPE to perform ping-pong reading and writing operations for lower delay.
It also has the function of fetching latest data address from feature extracting domain.

\textit{CtrlRf} is a dual port register file between control domain and VPE. 
Before computing, control domain writes configuration into ctrlRf, then start loading instruction and parameter. 
VPE will read configuration information from ctrlRf and wait for start signal.
After finishing computing tasks, ctrlRf generates FIN signal to control domain and wait for next round of computing.

\textit{VLIW} has four fields in an instruction word, each binds to one of the function modules. 
VLIW design enables programmer to issue the multiple instructions at the same time, maximizing the parallelism of VPE architecture.
Core instructions are listed in table \ref{VLIW}.
$prd$ and $prds$ are vector product instructions in SIMDU. $prd$ calls a eight-element-width vector product, while $prds$ calls two four-element-width ones. 
The data source of vector product is dRf, and the destination can be both dRf and address in on-chip memory. 
$vadd, vem$ in VU field stand for vector adding and vector element-wise multiplication. 
In Mif field, $fa$ is the abbreviation of fetching address from feature extracting domain, and $ld$ loads data from on-chip memory into dRf. 
$fin$ sets FIN signal to controlling field. 
Using instructions above, user can easily define general vector accelerating process for DL models.

\begin{table}
  \caption{Core VLIW instructions in VPE}
  \label{VLIW}
  \begin{tabular}{cccc}
    \toprule
    SIMDU field & VU field & Mif field & CtrlRf field\\
    \midrule
    $prd, dRf, dRf/adRf$ & $vadd, dRf, dRf/adRf$ & $fa$ & $fin$ \\
    $prds, dRf, dRf/adRf$ & $vem, dRf, dRf/adRf$ & $ld, adRf, dRf$ &  \\
  \bottomrule
\end{tabular}
\end{table}

\subsubsection{AryPE}
\begin{figure}[h]
  \centering
  \includegraphics[width=\linewidth]{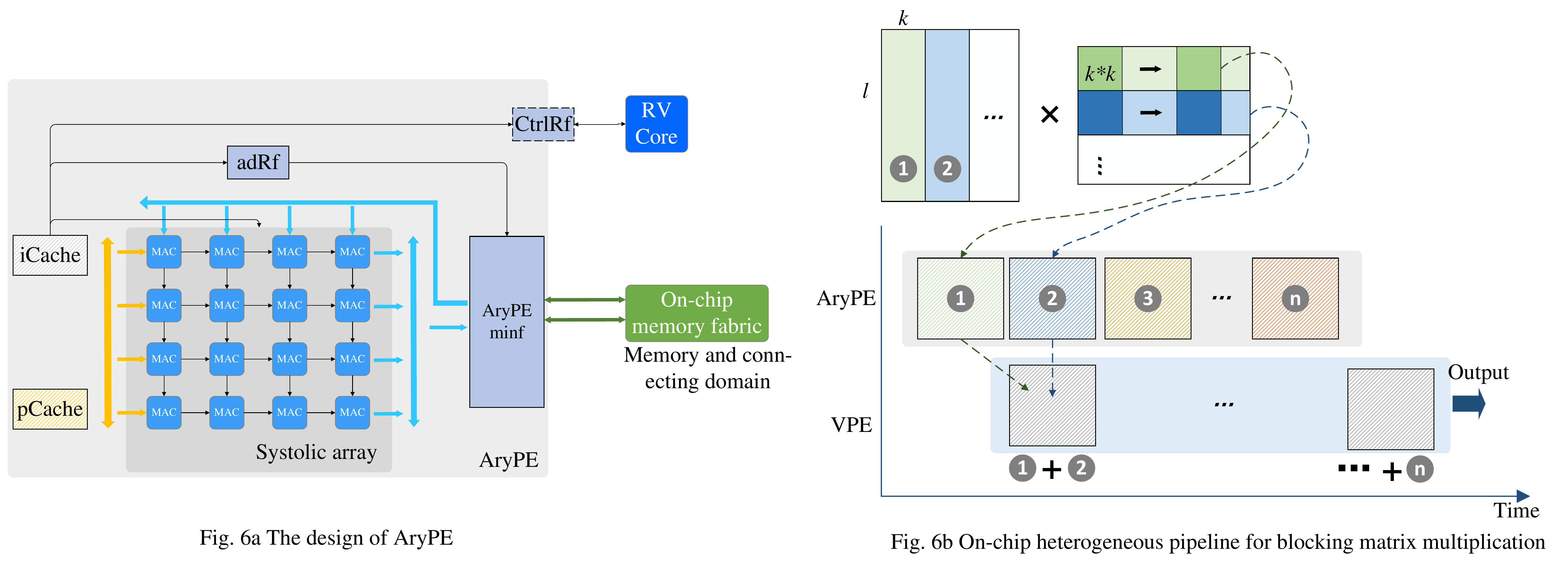}
  \caption{The design of AryPE and heterogeneous collaborative computing}
  \Description{Overview}
  \label{AryPE}
\end{figure}
AryPE is the high-throughput computing element for flow-based tasks in computing domain.
Like the structure of VPE, AryPE has an iCache for instructions, a pCache for parameters, an adRf for data address, a CtrlRf for receiving control information and a multi-channel Mif for high-efficiency memory access.

Different from VPE, the key computing module of AryPE is a $k \times k$ systolic array. 
It is proven that an ideal systolic array structure can achieve optimal throughput when conducting matrix multiplication. 
A matrix multiplication of $(l, k) \times (k, k)$ can be computed on a $k \times k$ systolic array without blocking. 
Core instructions of AryPE are quite simple: $(MM, l, \$x, \$y)$, $MM$ instruction represents matrix multiplication,  
$l$ is the length of streaming data, $\$x, \$y$ is different address registers in adRf, used for reading and writing data for systolic array;  
$(LD, \$p)$ instruction is for loading DL parameters from pCache, $\$p$ is an address register in adRf.


As mentioned before, systolic array faces two problems in actual DL accelerating: under-utilization and matrix blocking. 
Two problems bring computing efficiency loss to AryPE.
To address problems, heterogeneous collaborative computing between VPE, AryPE is designed.


\subsubsection{Heterogeneous collaborative computing}
Heterogeneous collaborative computing works between VPE, AryPE and on-chip memory fabric.
We take the first two convolution layers of CNN model used in \cite{zeroday} as an example to show how heterogeneous collaborative computing works.
A convolution layer with $w$ sliding windows, $ic$ input feature channels, $oc$ output feature channels and $s$ convolution kernel size can be mapped to a matrix multiplication of $(w, ic \times s) \times (ic \times s, oc)$ via img2col function.
The dimension of first layer feature in example is $(1, 10)$, there are 32 convolution kernels with kernel size of 3. Strides is 1, so the number of sliding windows is 10.
Finally, the computation of first layer is mapped to a matrix multiplication of $(10, 3) \times (3, 32)$. 
The input feature of second layer is of $(10, 32)$, so $ic = 32$. There are 32 convolution kernels with kernel size of 3, strides is 1, the number of sliding windows is 10. The computation of second layer is mapped to a matrix multiplication of $(10, 96) \times (96, 32)$. 
We take $k = 32$, getting a systolic array of $32 \times 32$.
It is obvious that the first layer introduces under-utilization problem for only achieving 9.3\% utilization of computing units, 
the second layer brings blocking problems for its computing scale exceeding the size of systolic array.

To address under-utilization, we offload the computing of first layer to VPE, using flexible SIMDU to conduct three-element-width vector product of first layer. Its convolution kernel size can fit in sub-lanes efficiently. 
The output of SIMDU is sent to the shared space in on-chip memory fabric, and then fetched by AryPE for following computing. 
In this case, on-chip heterogeneous pipeline offloads small-scale matrix multiplication to VPE, saving computing efficiency for AryPE. 

The second convolution layer faces blocking matrix problem. 
To run on a $32 \times 32$ systolic array, original matrix multiplication of $(10, 96) \times (96, 32)$ has to be split into nine sub-operations of $(10, 32) \times (32, 32)$ and eight aggregations of $(10, 32) + (10, 32)$.
VU in VPE is used for aggregations in the pipeline, as shown in figure \ref{AryPE}b. 
Ping-pong buffer is assigned in on-chip memory fabric between AryPE and VU.
When a temporal block is computed by AryPE, it will be written into the buffer. 
VU fetches the block, aggregates it with previous blocks, and sends aggregations result back to the memory fabric.
With this overlap collaborating, AryPE and VPE work in parallel. AryPE need not to stop matrix multiplication for block aggregating, as it being offloaded to VU.
The case of CNN illustrates the function and benefit of heterogeneous collaborating, 
where hardware efficiency and computing performance is both ensured.

\subsection{Memory and connecting domain}

Memory and connecting domain contains the on-chip memory fabric.
On-chip memory fabric has two components: 
(i) feature memory is for storing flow or packet feature;
(ii) computing memory is for data exchanging between heterogeneous modules, storing DL parameter, intermediate data and final computing results.

The data source of feature memory is from feature extractor, and the output destination are VPE and AryPE. 
In order to support simultaneously access of two domains, feature memory is implemented by independent memory bank with true dual ports.
As discussed before, computing memory provides storage and data exchanging for computation of DL models.
Heterogeneous collaborative computing depends on computing memory for transferring intermediate data between VPE and AryPE.
To support concurrent access and ping-pong buffer for two computing components, computing memory contains multi-bank true dual ports memory elements, each bank is connected to one memory channel in VPE and AryPE.
Computing memory can be accessed by control domain as well.
Control domain needs to read inference results of DL models to generate flow tables.
Before start computing, read-only parameters of DL models is also written into the computing memory by control domain from on-board flash or other offline storage element.

\subsection{Control domain}
Control domain has three functions: 
(i) loading initial instructions and parameter of DL models for computing domain;
(ii) sending configurations and controlling information to function modules;
(iii) transforming inference result of DL models into traffic rule-tables and updating data-plane.

Octopus takes one or multiple RV cores as controller in the domain. 
In initializing stage, RV core moves read-only DL parameter and programmed instructions from peripheral to on-chip memory fabric, and then loads them into function modules. 
Groups of control register files is set between control domain and other function modules. 
Controlling information is configured by the register writing from RV core. 
When finishing the computation of DL models, computing domain will send the address of output data and FIN signal to control register files. 
RV core keeps monitoring the register file, fetches inference results when triggered, parses results and finally generates decision actions to data-plane.
The format of decision actions can be rule-tables or controlling packets, depending on specific data-plane. 
The programmability of RV core provides flexibility for different controlling cases.

\section{Implementation and evaluation}
\subsection{Hardware implementation}
We implement the accelerator design on FPGA to further verify Octopus architecture. 
In detail, we design an 8k-depth flow-state table in flow tracker, enabling feature extractor to track up to 8k different flows. 
In memory and connecting domain, one true-dual-port Block RAM (BRAM) of 8k depth and 128 bits width is arranged as feature memory. Two true-dual-port BRAMs with 16k depth and 128 bits width are computing memory. Each of them binds to a memory access channel in VPE and AryPE. 
So modules in computing domain has two memory channels, which supports on-chip heterogeneous collaborative computing effectively.
In computing domain, we install eight lanes for a SIMDU, and eight parallel adders/multiplication units for VU in VPE.
A $16 \times 16$ size of systolic array is implemented in AryPE. 
Such organizing scheme of computing units can fully utilize the bandwidth of dual memory channels. 
To further increase hardware efficiency, all computing units in Octopus are for operations of Int-8 format, which has been proved will not influence the accuracy of DL models greatly \cite{tarurs, me, n3ic}. 
As to control domain, one open-source PULP \cite{spin, pspin} RV core is integrated, with SPI and CAN interface for interactions between accelerator and peripheral equipment. 

Octopus is implemented in Verilog HDL, evaluation platform is Xilinx xcku-115 FPGA and Vivado 2018.3 IDE.
Hardware resource occupation and frequency are listed in table \ref{imp result}, and "pct" stands for the occupation percentage of on-board resource.
Feature extractor in our design keeps a good balance between fast extracting and resource utilization. 
It occupies less than 1.5\% of available resource on FPGA and achieves 125MHz of frequency, providing an extracting throughput more than 31Mpkt/s. 
Assuming average packet size is 500 byte, our feature extractor can match over 124Gbps throughput of data-plane.

VPE and AryPE occupies 402 DSP units for computing, providing computing power of 145GOP/s, with computing memory banks in on-chip memory fabric offering 256KB storage space only under 4.7\% BRAM utilization. 
This leaves sufficient scaling space for more powerful computing ability. 
PULP RV core achieves 45Mhz frequency on FPGA platform, which is mainly restricted by un-optimized branch functions. However, RV core mainly competes initialization and controlling functions, which will not influence main computing process. 

The implementation results fully show the efficiency of our feature extractor, the performance of VPE and AryPE, and the scaling potential for higher performance. 

\begin{table}
  \caption{Implementation results}
  \label{imp result}
  \begin{tabular}{ccccc}
    \toprule
    Module&LUT/pct&BRAM/pct&DSP/pct&Frequency(Hz)\\
    \midrule
    Feature extractor&9051/1.4\%&21.5/1.0\%&0/0&125M\\
    On-chip memory fabric&623/0.1\%&128.5/5.9\%&0/0&125M/222M\\
    VPE&3153/0.5\%&17/0.8\%&141/2.6\%&222M\\
    AryPE&11000/1.7\%&26.5/1.2\%&256/4.6\%&222M\\
    RV core&11634/1.8\%&37/1.7\%&0/0&45M\\
    Total&35451/5.3\%&230.5/10.7\%&402/7.3\%&-\\
  \bottomrule
\end{tabular}
\end{table}

\subsection{Micro-benchmarks}
To fully validate the functionality and performance of Octopus, we organize three real use-cases for deploying DL for network models. 
The chosen use-cases contain both packet and flow-based granularity, take different traffic feature as input and utilize different DL models.
We believe these use-cases reflect typical DL for network models in real world.
The performance is acquired from cycle-accurate register-transfer-level hardware simulator and Octopus implementation results. 
To be noticed, we focus on validating the Tarurs ability of enabling DL models, thus accuracy and performance of specific DL models are out of the content in this paper.

\textit{\textbf{Use-case 1: a packet-based MLP model using packet size for intrusion detection}} \cite{usecase1}. 
Tarurs takes this use-case as well, which can clearly compare the performance of two architectures. 
The input feature of the MLP model is a six-dimension vector, containing packet size, packet direction and other features for each packet. 
Used features are easily to acquired via our feature extractor.
The output of MLP is a binary prediction of whether it is a malicious traffic.
The size hidden layer of the model is 12, 6, 3, 2, represented by matrix of (6, 12), (12, 6), (6, 3) and (3, 2) dimension respectively. 

\begin{figure}[h]
  \centering
  \includegraphics[width=0.7\linewidth]{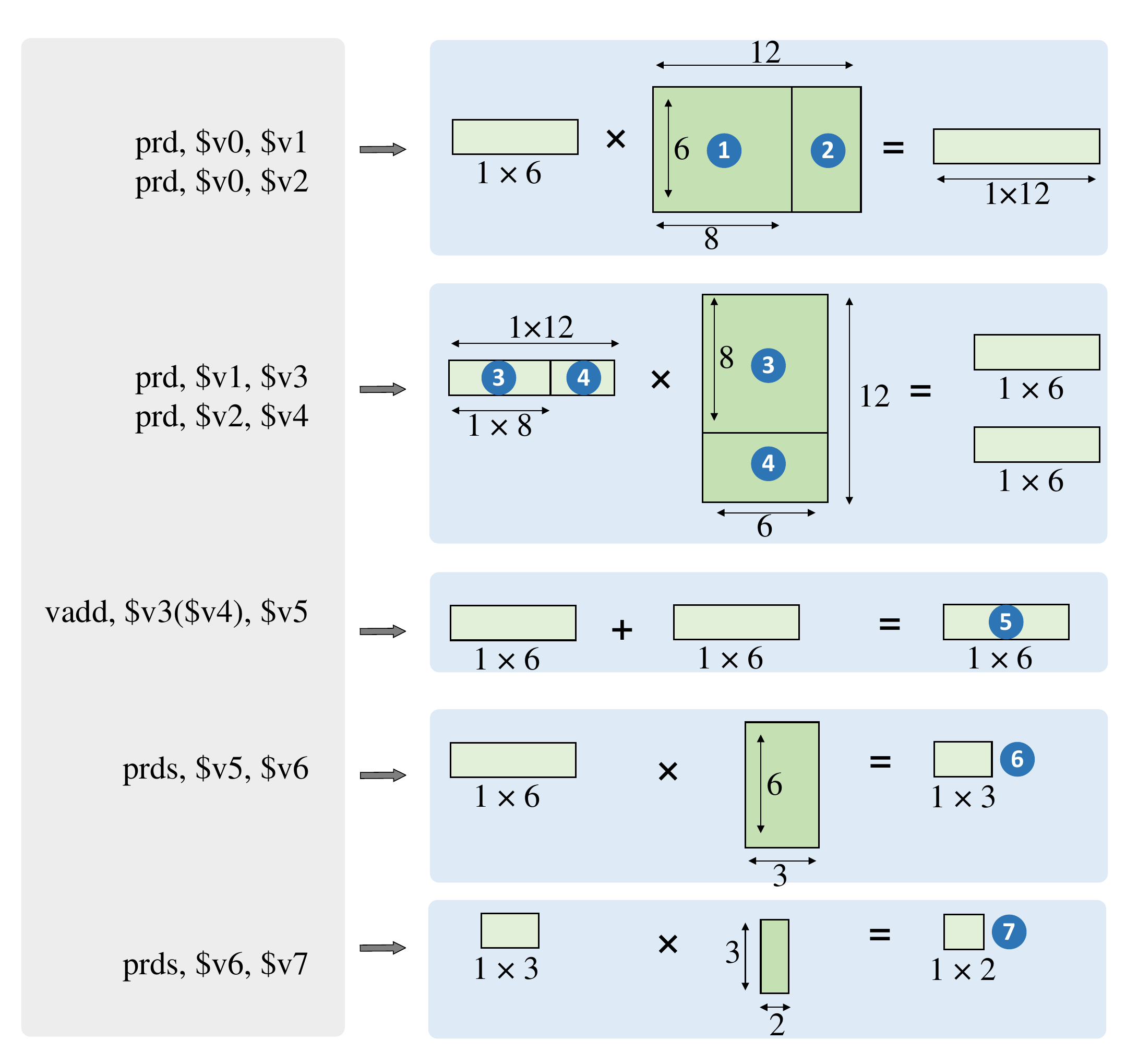}
  \caption{Computing kernel of the packet-based MLP model}
  \Description{accl_arch}
  \label{usecase2}
\end{figure}

In this use-case, packet-based computation task is allocated to VPE. 
The kernel instructions of computing is listed in figure \ref{usecase2}. 
Four \textit{prd} instructions perform vector productions between input feature and DL parameter in order, computing first two layers in MLP (\textcircled{1}-\textcircled{4}). 
Parameter is implicitly sent to SIMDU during decoding stage for per instruction.
The relative small computing dimension is well-suited in the hierarchy of lane and sub-lane. 
In step \textcircled{5}, vector adding (\textit{vadd}) is needed to accumulate temporal vectors. 
Considering the small size of last two layers, sub-lane operations are applied via \textit{prds} instruction to compute final output of MLP model(\textcircled{6}, \textcircled{7}).
With 222Mhz frequency, the feature extracting and computing delay of this packet-based model is 207ns, which matches the latency of packet forwarding and out-performs Tarurs. 

To deeply compare the design and performance difference reflected in this use-case, we list detail information of Octopus and Tarurs in table \ref{tarurs compare}.
Although Tarurs and Octopus have similar computing delay and hardware scale, there exists important difference in two architectures. 
Tarurs possesses higher computing frequency, but their pipeline architecture consumes 16 computing units (FU) per stages, leading to low computing efficiency and insufficient scalability. 
On contrast, Octopus adopts vector accelerator, applying SIMD and VLIW structure to ensure the efficiency of parallelism and sufficient scalability . 
Also, flexible hierarchy structure further ensures computing efficiency for DL models.

\begin{table}
  \caption{Comparsion between Octopus and Tarurs under usa-case1}
  \label{tarurs compare}
  \begin{tabular}{ccccc}
    \toprule
    Works &Architecture&Delay&Scale & Frequency\\
    \midrule
    Tarurs&pipeline&221 ns& 4 stages $\times$ 16 FU&1Ghz\\
    Octopus&vector accelerator&207 ns& 8 $\times$ 2 sub-lanes &222Mhz\\
  \bottomrule
\end{tabular}
\end{table}

\textit{\textbf{Use-case 2: a flow-based CNN model using arriving interval time for traffic classification}} \cite{zeroday}.
This flow-based model is an 1D-CNN as pictured in figure \ref{dl model}b, which is a common choice in the field \cite{zeroday, survey0}.
Such 1D-CNN model takes the vector of top-20 packets arriving interval time of a flow as input and can be directly acquired by our feature extractor.
There are three convolution layers in the model, with size of \{kernel\_size=3, input\_channel=1, output\_channel=32\}, 
\{kernel\_size=3, input\_channel=32, output\_channel=32\},
\{kernel\_size=3, input\_channel=32, output\_channel=32\} for each. 
Max-pooling layers with stride of 2 is between every convolution layer. 
After convolution and pooling layers are a fully-connect (FC) layer with 128 output units and a linear layer with 162 output units. 
The output contains the fine-grained classification results of traffic from 162 different applications. 
The computations of three convolution layers, FC layer and linear layer are mapped to the matrix multiplication of $[(20 \times f, 3) \times (3, 32)]$, $[(10 \times f, 96) \times (96 \times f, 32)]$, $[(5 \times f, 96) \times (96, 32)]$, $[(f, 96) \times (96, 128)]$, $[(f, 128) \times (128, 162)]$, respectively. $f$ represents the number of tracked flows. 

The computation of first CNN layer brings under-utilization problem for AryPE, while other layers bring blocking matrix multiplication issue. 
As mentioned above, we apply heterogeneous collaborative computing for this use-case. 
The matrix multiplication of $[(20 \times f, 3) \times (3, 32)]$ in first layer is offloaded to SIMDU in VPE, AryPE focuses on large matrix multiplication in deeper layers, and blocking matrix aggregation is allocated to VU in VPE. 
We set $f=1000$ and conduct simulation with and without heterogeneous collaborative computing. Results are shown in table \ref{case2}.
Experiment results clearly show that (i) our heterogeneous collaborative computing can increase computing efficiency and performance, resulting in 1.69$\times$ speedup on throughput; (i) straightforwardly adding existing accelerators into data-plane may not bring expected performance. 

\begin{table}
  \caption{Performance of heterogeneous collaborative computing}
  \label{case2}
  \begin{tabular}{ccccc}
    \toprule
    Method &SIMDU.eff&VU.eff&AryPE.eff & Throughput\\
    \midrule
    w/ collaborating&12.1\%&83.8\%& 81.1\%&90kflow/s\\
    wo/ collaborating&-&-& 48.2\% &53kflow/s\\
    
  \bottomrule
\end{tabular}
\end{table}

\textit{\textbf{Use-case 3: a flow-based transformer model using payload for traffic classification}} \cite{usecase3}.
The transformer structure of use-case is depicted in figure \ref{dl model}.
\cite{usecase3} designs several formats for input feature of transformer, we choose the following option: the input is a payload matrix from top-16 bytes of top-15 packets in a flow. 
In self-attention module, the size of \textit{WQ}, \textit{WQ}, \textit{WQ} parameter matrix is $(16, 64)$,
so the main computation in figure \ref{dl model}c are three $[(15, 16) \times (16, 64)]$ matrix multiplication, a $[(15, 64) \times (64, 15)]$ one, a $[(15, 15) \times (15, 64)]$ one. Following self-attention module is a MLP layer, with input size of 64, hidden layer size of 128 and output layer size of 64, as mentioned in figure \ref{dl model}a. 
The computation of MLP here is matrix multiplication of $[(15, 64) \times (64, 128)]$ and $[(15, 128) \times (128, 64)]$. 

The computation of transformer models is larger than CNN and MLP, which may impede their application in network tasks. 
However, Octopus shows potential for enabling transformer models in experiments. 
The main challenge in this use-case is large-scale block matrix multiplication, so we apply heterogeneous collaborative computing to offload block matrix aggregations. 
The number of tracked flows is also set as $f=1000$, each flow is sent to one-stage 
self-attention modules and MLP in transformer model.
In this case, AryPE achieves computing efficiency of 96.3\% and throughput of 35.7 kflow/s. 

To conclude, we choose three use-cases to validate both functionality and performance of Octopus accelerator.
These use-cases include packet-based and flow-based task, small and large size computation, under-utilization and block matrix multiplication cases, MLP, CNN and transformer models. 
Experiment results show for packet-based tasks, Octopus can offer comparable computing latency with packet forwarding; 
for flow-based tasks, Octopus can achieve 35.7kflow/s to 90kflow/s throughput with collaborative computing.

\section{Related works}
There are several works on deploying machine-learning-based network models like decision tree and K-means on P4 hardware \cite{iisy, dream, henna}. 
IIsy \cite{iisy} and \cite{henna} deploy decision tree and random forests models on P4 device for real-time traffic classification, 
\cite{dream} run k-means clustering models on P4 hardware. 
However, as mentioned before, P4 device lacks computing power to deploy a real DL-based model, and the computing complexity of such machine-learning models is far lower than DL models.
Similarly, many-core based data-plane also lacks computing power for DL models, and it sometimes is not equipped with hardware feature extractor. Instead, it depends on CPU programs to extract feature, which is time-consuming and low-efficiency.

There are few works researching on implementing DL models in data-plane \cite{n3ic, tarurs,fenbushi}.
Tarurs\cite{tarurs} is the most related work to ours. 
Tarurs simply inserts a specific CNN hardware engine between ingress and engress pipelines of a P4 device. 
The DL inference delay of Tarurs is around hundreds of ns, meeting the delay requirement of packet-based tasks.
However, Tarurs only focuses on per-packet tasks, ignoring the flow granularity. 
Besides that, the accelerating engine of Tarurs is designed with fixed pipeline stages, fails to realize generality. 
This structure significantly limits the scale and choice of DL for network models running on their device.
Lastly, Tarurs relies on P4 hardware to provides fast feature extracting. They do not design their own feature extractor.
Apart from Tarurs, N3IC \cite{n3ic} proposes to deploy a binary neural network (BNN) on in-network hardware. 
Their BNN model transforms all feature and parameter into a binary bit, then complex computation is replaced by bit-wise operations like \textit{and} and \textit{or}.
Although BNN is hardware-friendly, it bears non-negligible accuracy loss, and N3IC still lacks necessary computing power for DL models.
\cite{fenbushi} proposes to split a DNN model on several data-plane devices in a network path. 
However, this work does not improve the architecture of data-plane, and distributing method is hard to be practicable in real network environment.

\section{Conclusion}
In this paper, we propose Octopus: a heterogeneous in-network computing accelerator enabling DL for network. 
We first clarify four key challenges for data-plane to support DL-based network models, then we design Octopus to address above problems.
The proposed architecture includes feature extractor, VPE, AryPE, on-chip memory fabric and RV core.
A heterogeneous collaborative computing mode is also designed to further increase performance. 

To fully validate the performance and functionality of Octopus, 
we implement the proposed architecture on FPGA and four use-cases are arranged.
In experiments, Octopus possesses 31Mpkt/s feature extracting ability with lower hardware resource occupation, provides 207 ns packet-based computing latency and up to 90kflow/s flow-based inference throughput. 
Experiments shows that four key demands of computing power, traffic granularity, model generality and feature extracting are sufficiently satisfied by Octopus.

\bibliographystyle{ACM-Reference-Format}
\bibliography{sample-base}

\appendix
\begin{table*}
  \caption{Appendix: Whole feature set}
  \begin{tabular}{lll}
    \toprule
    No. &Feature &Category\\
    \midrule
    1 & payload size & size\\ 
    2 & payload & payload\\ 
    3 & packet arrival interval time & time\\ 
    4 & packet direction & direction\\ 
    5 & IP tuple & protocol\\ 
    6 & flow size & size\\ 
    7 & flow size with two directions & size, direction\\ 
    8 & protocol type & protocol\\ 
    9 & flow duration time & time\\ 
    10 & flow duration time with two directions & time, direction\\
    11 & max packet length & size\\
    12 & min packet length & size\\
    13 & mean packet length & size\\
    14 & variance of packet length & size\\
    15 & max packet length with two directions& size, direction\\
    16 & min packet length with two directions& size, direction\\
    17 & mean packet length with two directions& size, direction\\
    18 & variance of packet length with two directions& size, direction\\
    19 & max packet arrival interval time & time\\
    20 & min packet arrival interval time & time\\
    21 & mean packet arrival interval time & time\\
    22 & variance of packet arrival interval time & time\\
    23 & max packet arrival interval time with two directions& time, direction\\
    24 & min packet arrival interval time with two directions& time, direction\\
    25 & mean packet arrival interval time with two directions& time, direction\\
    26 & variance of packet arrival interval time with two directions& time, direction\\
    27 & TCP window size & protocol\\
    28 & TCP/UDP flag & protocol\\
    30 & packet per second & size, time\\
    31 & bytes per second & size, time\\
    32 & packet per second with two directions& size, time, direction\\
    33 & bytes per second with two directions& size, time, direction\\
    34 & packet per second on the port & size, time, direction\\
    35 & bytes per second on the port & size, time, direction\\
    36 & total number of packet & size\\
    37 & total number of packet with two directions& size, direction\\
    38 & vector of packet size & size\\
    39 & vector of packet size with two directions& size, direction\\
    40 & vector of packet arrival interval time & time\\
    41 & vector of packet arrival interval time with two directions & time, direction\\
  \bottomrule
\end{tabular}
\end{table*}

\end{document}